\documentclass[12pt, a4paper]{article}
\usepackage{graphicx}
\usepackage{epsfig}
\usepackage{jcappub}
\usepackage{epstopdf}
\def \omm  {\Omega_{0 {\rm m}}}

\begin{document}

\title{Cosmic acceleration from coupling of known components of matter:
Analysis and diagnostics}
\author{Abhineet Agarwal$^{1}$, R. Myrzakulov$^{2}$, S. K. J. Pacif$^{3}$, M. Shahalam$^{4}$}

\affiliation{$^1,^3$Centre for Theoretical Physics, Jamia Millia Islamia, New Delhi 110025, India}
\affiliation{$^2$Eurasian International Center for Theoretical Physics and Department of General
\& Theoretical Physics, Eurasian National University, Astana 010008, Kazakhstan}
\affiliation{$^4$Institute for Advanced Physics \& Mathematics, Zhejiang University of Technology, Hangzhou 310032, China}
\emailAdd {abuwdu123@gmail.com, rmyrzakulov@gmail.com, shibesh.math@gmail.com, shahalam@zjut.edu.cn}
\abstract{ In this paper, we examine a scenario in which late-time cosmic acceleration might arise due to the coupling between baryonic matter and dark matter without the presence of extra degrees of freedom. In this case, one can obtain late-time  acceleration in Jordan frame and not in Einstein frame. We consider two different forms  of parametrization of the coupling function, and put constraints on the model parameters by using an integrated datasets of Hubble parameter, Type Ia supernova and baryon acoustic oscillations. The models under consideration are consistent with the observations. In addition, we perform the statefinder and $Om$ diagnostics, and show that the  models  exhibit a distinctive behavior due to the phantom characteristic in future which is a generic feature of the underlying scenario.}

\date{\today }
\maketitle

\section{Introduction}
\label{sect1}
Late-time cosmic acceleration is an inevitable ingredient of
our Universe directly supported by cosmological observations \cite{HZTEAM,
SCP}. Other observations such as cosmic microwave background (CMB),
baryonic aucostic oscillations (BAO), sloan digital sky survey \cite{cmb1, cmb2,
bao1, bao2, sdss1, sdss2} and many others support this fact indirectly. The
globular cluster reveals that the age of certain objects in the
Universe is larger than the age of Universe estimated in standard model with
normal matter. The only known resolution of the puzzle is provided by
invoking cosmic acceleration at late-times. Although, there are many ways to
explain the accelerating expansion of the Universe (e.g. by adding a source
term in the matter part of Einstein field equations, by modifying the
geometry or by invoking inhomogeneity), the inclusion of a source term with
large negative pressure dubbed ``\textit{dark energy}" (DE) \cite{DE1, DE2, DE3,
DE4, DE5, DE6, DE7, DE8, DE9, DE10} is widely accepted to the theorists.
However, a promising candidate for dark energy (cosmological constant $\Lambda $) is
under scrutiny. A wide varieties of dark energy candidate have been proposed
in the past few years such as cosmological constant \cite{DE3, carollCC,
padmaCC, peeblesCC}, slowly rolling scalar field \cite{quint1, quint2,
quint3, quint4, quint5}, phantom field \cite{phant1, phant2, phant3, phant4,
phant5, phant6}, tachyon field \cite{tachy1, tachy2,alam2017} and chaplygin gas \cite%
{CG1, CG2} etc. (See \cite{DE1, DEREV1, DEREV2, DEREV3} for a detailed list).

The modifications of Einstein's theory of gravity not only account the cosmic acceleration but also resolves many standard problems
such as singularity problem, the hierarchy problem, quantization and
unification with other theories of fundamental interactions. Massive
gravity, Gauss-Bonnet gravity, $f(R)$, $f(T)$, $f(R,T)$ gravities,
Chern-Simon gravity, Galileon gravity are name a few among the various
alternative theories proposed in the past few years. Modified gravity can
also provide unified description of the early-time inflation with that of
late-time cosmic acceleration and dark matter (DM). Traditionally, all these
modifications invoke extra degrees of freedom non-minimally coupled to
matter in the Einstein frame. Generally, it is believed that late-time
acceleration requires the presence of dark energy or the extra degrees of
freedom. Recently, Berezhiani et al. \cite{KHOURY2016} discussed a third
possibility which requires neither any exotic matter nor the large scale
modifications of gravity. They showed that the interaction between the
normal matter components, namely, the dark matter and the baryonic matter (BM) 
can also provide the late-time acceleration in Jordan frame. In context of the
coupling, the stability criteria disfavors the conformal coupling while the
maximally disformal coupling can give rise to late-time cosmic acceleration
in Jordan frame but no acceleration in the Einstein frame. Extending the
work of Ref. \cite{KHOURY2016}, Agarwal et al. \cite{ARSSA} have further
investigated the cosmological dynamics of the model obtained by
parameterizing the coupling function. Also, they have shown that the model
exhibits the sudden future singularity which can be resolved by taking a
more generalized parametrization of the coupling function.

In this paper, we shall consider two forms of parametrizations. The first parametrization shows the sudden future singularity, which can be pushed into far future if we consider the second parametrization. We further investigate the models using statefinder and $Om$ diagnostics. The paper is organized as follows. Section \ref{sect:2} is devoted to the basic equations of the models. In section \ref{sec:obs}, we put the observational constraints on the model parameters. The detailed analysis of statefinder and $Om$ diagnostics are presented in sections \ref{sec:state} and \ref{sec:om}, respectively. We conclude our results in section \ref{sec:conc}.
\section{Field equations}
\label{sect:2}
The scenario of the interaction between dark matter and the baryonic
matter as described briefly in Refs. \cite{KHOURY2016} and \cite{ARSSA} in a spatially
flat Freidmann-Lemaitre-Robartson-Walker (FLRW) background
\begin{equation}
ds^{2}=-dt^{2}+a^{2}(t)\left( dx^{2}+dy^{2}+dz^{2}\right) \text{,}  \label{1}
\end{equation}%
yield the field equations,
\begin{equation}
3H^{2}=8\pi G\left( \Lambda _{DM}^{4}\sqrt{\frac{X}{X_{eq}}}\left( \frac{%
a_{eq}}{a}\right) ^{3}-P+QR^{3}\tilde{\rho}_{b}\right) \text{,}  \label{2}
\end{equation}%
and
\begin{equation}
2\frac{\ddot{a}}{a}+H^{2}=-8\pi G(P+P_{b})\text{,}  \label{3}
\end{equation}%
where $Q$ and $R$ are two arbitrary coupling functions.

The Einstein frame metric couples to Jordan frame metric such that $\sqrt{-%
\tilde{g}}=QR^{3}\sqrt{-g}$ ($\tilde{g}$ is the determinant of Jordan frame
metric $\tilde{g}_{\mu \nu }$, constructed from the Einstein frame metric $%
g_{\mu \nu }$) and $X=-g^{\mu \nu }\partial _{\mu }\Theta \partial _{\nu
}\Theta $, $\Theta $ being the dark matter field. The quantities $P$ and $P_{b}$ are
pressures of DM and BM in Einstein frame respectively which are related by $%
P_{b}\equiv QR^{3}\tilde{P}_{b}$. $\tilde{P}_{b}$ and $\tilde{\rho}_{b}$ are
the pressure and density of BM in Jordan frame and are related to BM density
in Einstein frame by the relation $\rho _{b}=QR^{3}\left( \tilde{\rho}%
_{b}\left( 1-2X\frac{Q,_{X}}{Q}\right) +6X\frac{R,_{X}}{R}\tilde{P}%
_{b}\right) $. For the detailed derivation of field equations, see \cite%
{KHOURY2016} and \cite{ARSSA}. Here, we shall note that all the quantity with a
tilde above are in Jordan frame and without tilde are in Einstein frame.

The Jordan frame scale factor dubbed \textit{physical scale factor} is
related a scale factor of Einstein frame  as
\begin{equation}
\tilde{a}=Ra\text{.}  \label{4}
\end{equation}
\qquad Here, we consider the same maximally disformal coupling of BM and DM for which $Q=1$
throughout the evolution and $R=1$ in the early Universe that grows
sufficiently fast such that the physical scale factor $\tilde{a}$ in Jordan
frame experiences acceleration. The conformal coupling is disfavored by the stability criteria \cite{KHOURY2016}. One needs to specify the coupling function $%
R(a)$ to proceed further or equivalently, $a$ can be parametrized in terms
of physical scale factor $\tilde{a}$. The two parametrizations are
\\
(1) Model 1:  $a(\tilde{a})=\tilde{a}+\alpha \tilde{a}^{2}+\beta \tilde{a}^{3}$, where 
$\alpha $ and $\beta $ are two model parameters.
\\
(2) Model 2:  $a(\tilde{a})=\tilde{a}e^{\alpha \tilde{a}}$, in this case, only $\alpha $ is a model parameter.

By expanding the functional $\tilde{a}e^{\alpha \tilde{a}}$ in Taylor
series, the first parametrization $\tilde{a}+\alpha \tilde{a}^{2}+\beta 
\tilde{a}^{3}$ can be recovered by substituting $\beta=\alpha^2/2$. Agarwal et al. \cite{ARSSA} have studied
various features of the model 1 and constrained the parameters $\alpha $ \& $%
\beta $ by employing the $\chi ^{2}$ analysis using $H(z)+SN+BAO$
datasets. Extending the analysis, we further study some more physical
characteristics of the models 1 and 2 such as the statefinder and $Om$ diagnostics, and also put observational constraints on the parameter $\alpha $ of model 2.

We also need to express the cosmological parameters in terms of redshifts in
both the Einstein frame and Jordan frame which are defined as
\begin{equation}
\tilde{a}=\frac{\tilde{a}_{0}}{1+\tilde{z}}\text{ , \ \ }a=\frac{a_{0}}{1+z}%
\text{,}  \label{5}
\end{equation}%
For both the parametrizations, $\tilde{a}_{0}=1$ but $a_{0}=1+\alpha+\beta\neq 1$ (model 1) and $a_{0}=e^\alpha \neq 1$ (model 2).

For model 1, the explicit expressions for the Hubble and deceleration parameters in Jordan frame are obtained as
\begin{equation}
\tilde{H}(\tilde{a})=\frac{\dot{\tilde{a}}}{\tilde{a}}=\tilde{H}_{0}\frac{%
(1+\alpha +\beta )^{\frac{1}{2}}(1+2\alpha +3\beta )}{\tilde{a}^{\frac{3}{2}}%
\left[ 1+\alpha \tilde{a}+\beta \tilde{a}^{2}\right] ^{\frac{1}{2}}\left[
1+2\alpha \tilde{a}+3\beta \tilde{a}^{2}\right] }\text{,}  \label{eq:H1}
\end{equation}
and%
\begin{equation}
\tilde{q}(\tilde{a})=-\frac{\tilde{a}\ddot{\tilde{a}}}{\dot{\tilde{a}}^{2}}=%
\frac{1}{2}\left( \frac{1+2\alpha \tilde{a}+3\beta \tilde{a}^{2}}{1+\alpha 
\tilde{a}+\beta \tilde{a}^{2}}\right) +\left( \frac{2\tilde{a}(\alpha
+3\beta \tilde{a})}{1+2\alpha \tilde{a}+3\beta \tilde{a}^{2}}\right) \text{.}
\label{eq:q1}
\end{equation}%
Using Eq. (\ref{5}), above expressions can be written in terms of redshift $\tilde{z}$ as 
\begin{equation}
\tilde{H}(\tilde{z})=\tilde{H}_{0}\frac{(1+\alpha +\beta )^{\frac{1}{2}%
}(1+2\alpha +3\beta )\left( 1+\tilde{z}\right) ^{\frac{9}{2}}}{\left[ \left(
1+\tilde{z}\right) ^{2}+\alpha \left( 1+\tilde{z}\right) +\beta \right] ^{%
\frac{1}{2}}\left[ \left( 1+\tilde{z}\right) ^{2}+2\alpha \left( 1+\tilde{z}%
\right) +3\beta \right] }\text{,}  \label{a3}
\end{equation}%
and%
\begin{equation}
\tilde{q}(\tilde{z})=\frac{\left[ \left( 1+\tilde{z}\right) ^{2}+2\alpha
\left( 1+\tilde{z}\right) +3\beta \right] ^{2}+4\left[ \alpha \left( 1+%
\tilde{z}\right) +3\beta \right] \left[ \left( 1+\tilde{z}\right)
^{2}+\alpha \left( 1+\tilde{z}\right) +\beta \right] }{2\left[ \left( 1+%
\tilde{z}\right) ^{2}+\alpha \left( 1+\tilde{z}\right) +\beta \right] \left[
\left( 1+\tilde{z}\right) ^{2}+2\alpha \left( 1+\tilde{z}\right) +3\beta %
\right] }\text{,}  \label{a4}
\end{equation}%
together with the effective equation of state (EOS) parameter given by
\begin{equation}
\tilde{w}_{eff}(\tilde{z})=\frac{\alpha \left( 5+6\alpha +5\tilde{z}\right)
(1+\tilde{z})^{2}+\beta (14+23\alpha +14\tilde{z})(1+\tilde{z})+18\beta ^{2}%
}{3\{(1+\tilde{z})^{2}+\alpha (1+\tilde{z})+\beta \}\{(1+\tilde{z}%
)^{2}+2\alpha (1+\tilde{z})+3\beta \}}\text{.}  \label{a5}
\end{equation}
Similarly, for model 2, we obtain the expressions for the Hubble and deceleration parameters in Jordan-frame as%
\begin{equation}
\tilde{H}(\tilde{a})=\frac{\tilde{H}_{0}(1+\alpha )e^{\frac{3}{2}\alpha }}{%
\tilde{a}^{\frac{3}{2}}\left[ 1+\alpha \tilde{a}\right] ^{\frac{1}{2}}e^{%
\frac{3}{2}\alpha \tilde{a}}},  \label{eq:H2}
\end{equation}%
and%
\begin{equation}
\tilde{q}(a)=\frac{\left( 1+\alpha \tilde{a}\right) ^{2}+2\tilde{a}(2\alpha
+\alpha ^{2}\tilde{a})}{2\left( 1+\alpha \tilde{a}\right) }\text{.}
\label{eq:q2}
\end{equation}%
with the help of Eq. (\ref{5}), we obtain%
\begin{equation}
\tilde{H}(\tilde{z})=\frac{\tilde{H}_{0}(1+\alpha )e^{\frac{3}{2}\alpha
}\left( 1+\tilde{z}\right) ^{\frac{5}{2}}}{\left[ \left( 1+\tilde{z}\right)
+\alpha \right] e^{\frac{3}{2}\displaystyle\frac{\alpha }{(1+\tilde{z})}}}%
\text{,}  \label{b3}
\end{equation}%
and%
\begin{equation}
\tilde{q}(\tilde{z})=\frac{1+\tilde{z}^{2}+6\alpha +3\alpha ^{2}+\left(
2+6\alpha \right) \tilde{z}}{2\left( 1+\tilde{z}\right) \left( 1+\tilde{z}%
+\alpha \right) }\text{.}  \label{b4}
\end{equation}
The effective equation of state is then given by
\begin{equation}
\tilde{w}_{eff}(\tilde{z})=\frac{5\alpha (1+\tilde{z})+3\alpha ^{2}}{3(1+%
\tilde{z})\left[ (1+\tilde{z})+\alpha \right] }\text{ .}  \label{b5}
\end{equation}
In both the models, another parameter $H_0$ will come in the expressions of $H(z)$, see Eqs. (\ref{a3}) and (\ref{b3}). But, here we focus on the parameters of underlying parametrizations. The first parametrization (model 1) consist of two model parameters (i.e. $\alpha$ and $\beta$) while the second parametrization (model 2) consists of a single model parameter (i.e. $\alpha$). Now we are in position to put the observational constraints on the parameters of model 2 in the following section. 

Before proceeding to next section, we consider The DGP model as \cite{dgp}:
\begin{equation}
\label{eq:DGP}
\frac{H(z)}{H_0} = \left[ \left(\frac{1-\omm}{2}\right)+\sqrt{\omm (1+z)^3+
\left(\frac{1-\omm}{2}\right)^2} \right]\,\,
\end{equation}
where $H_0$ and $\omm$ are the present values of Hubble parameter and energy density parameter of matter.
\section{Observational constraints}
\label{sec:obs}
We have already mentioned that the model 1 consists of two parameters, namely, $\alpha$ and $\beta$ which were constrained in Ref. \cite{ARSSA}. In our analysis, we shall use  their best-fit values given as $\alpha =-0.102681$ \& $ \beta =-0.078347$. In this section, we put the constraints on parameters of model 2 by employing the same procedure as in \cite{ARSSA}.

One can use the total likelihood to constrain the parameters $\alpha $ and $H_0$ of model 2. The total likelihood function for a joint analysis can be defined as 
\begin{equation}
\mathcal{L}_{tot}(\alpha , H_0 )=e^{-\frac{\chi _{tot}^{2}(\alpha , H_0 )}{%
2}}\text{,~~~~ where }\chi _{\mathrm{tot}}^{2}=\chi _{\mathrm{Hub}}^{2}+\chi _{%
\mathrm{SN}}^{2}+\chi _{\mathrm{BAO}}^{2}\text{.}  \label{o1}
\end{equation}%
Here, $\chi _{\mathrm{Hub}}^{2}$ denotes the chi-square for the Hubble dataset, 
$\chi _{\mathrm{SN}}^{2}$ represents the Type Ia supernova and $\chi _{%
\mathrm{BAO}}^{2}$ corresponds to the BAO. By minimizing the $\chi
_{\mathrm{tot}}^{2}$ , we obtain the best-fit value of $\alpha $ and $H_0$. The
likelihood contours are standard i.e. confidence level at 1$\sigma $ and 2$%
\sigma $ are $2.3$ and $6.17$, respectively in the 2D plane.

First, we consider 28 data points of $H(z)$ used by Farooq and Ratra \cite{Farooq:2013hq} in the redshift range $0.07\leq z\leq 2.3$, and use $H_{0}=67.8\pm 0.9~Km/S/Mpc$ \cite{planck2015}. The $\chi ^{2}$, in this case, is defined as 
\begin{equation}
\chi _{\mathrm{Hub}}^{2}(\theta )=\sum_{i=1}^{29}\frac{\left[ h_{\mathrm{th}%
}(z_{i},\theta )-h_{\mathrm{obs}}(z_{i})\right] ^{2}}{\sigma _{h}(z_{i})^{2}}%
\,,  \label{o2}
\end{equation}%
where $h=H/H_{0}$ represents the normalized Hubble parameter, $h_{\mathrm{obs}}$ and $h_{\mathrm{th}}$ are the observed and theoretical values of normalized Hubble parameter and $\sigma _{h}=\left( 
\frac{\sigma _{H}}{H}+\frac{\sigma _{H_{0}}}{H_{0}}\right) h$. The quantities $\sigma _{H}$ and $\sigma _{H_{0}}$ designate the errors associated with $H$ and ${H_{0}}$,
respectively.

Second, we use $580$ data points from Union2.1 compilation data \cite{Suzuki:2011hu}. The corresponding $\chi ^{2}$ is given as 
\begin{equation}
\chi _{\mathrm{SN}}^{2}(\mu _{0},\theta )=\sum_{i=1}^{580}\frac{\left[ \mu
_{th}(z_{i},\mu _{0},\theta )-\mu _{obs}(z_{i})\right] ^{2}}{\sigma _{\mu
}(z_{i})^{2}}\,,  \label{o3}
\end{equation}%
where $\mu _{obs}$, $\mu _{th}$ are the observed, theoretical distance
modulus and $\sigma _{\mu }$ is the uncertainty in the distance modulus, and $\theta $ is an arbitrary parameter. The distance modulus $\mu (z)$ is an observed quantity and related to luminosity distance $D_{L}(z)=(1+z)\int_{0}^{z}\frac{H_{0}dz^{\prime }}{%
H(z^{\prime })}$ as $\mu (z)=m-M=5\log D_{L}(z)+\mu _{0}$; $m$ and $M$ being the apparent and absolute magnitudes of the supernovae, and $\mu _{0}=5\log
\left( \frac{H_{0}^{-1}}{\mathrm{Mpc}}\right) +25$ is a nuisance parameter

Finally, we consider BAO data. The corresponding chi-square ($\chi _{\mathrm{BAO}}^{2}$) is defined by 
\cite{Giostri:2012ek}: 
\begin{equation}
\chi _{\mathrm{BAO}}^{2}=Y^{T}C^{-1}Y\,,  \label{o4}
\end{equation}%
where
\begin{equation}
Y=\left( \begin{array}{c}
        \frac{d_A(z_\star)}{D_V(0.106)} - 30.95 \\
        \frac{d_A(z_\star)}{D_V(0.2)} - 17.55 \\
        \frac{d_A(z_\star)}{D_V(0.35)} - 10.11 \\
        \frac{d_A(z_\star)}{D_V(0.44)} - 8.44 \\
        \frac{d_A(z_\star)}{D_V(0.6)} - 6.69 \\
        \frac{d_A(z_\star)}{D_V(0.73)} - 5.45
        \end{array} \right)\,,
\end{equation}
and inverse covariance matrix ($C^{-1}$), values $\frac{%
d_{A}(z_{\star })}{D_{V}(Z_{BAO})}$ are taken into account as in \cite%
{Blake:2011en, Percival:2009xn, Beutler:2011hx, Jarosik:2010iu,
Eisenstein:2005su, Giostri:2012ek}, and $z_{\star }\approx 1091$ is the
decoupling time, $d_{A}(z)$ is the co-moving angular-diameter distance and $%
D_{V}(z)=\left( d_{A}(z)^{2}z/H(z)\right) ^{1/3}$ is the dilation scale.

\begin{figure}[tbph]
\begin{center}
{\includegraphics[width=2.3in,height=2.3in,angle=0]{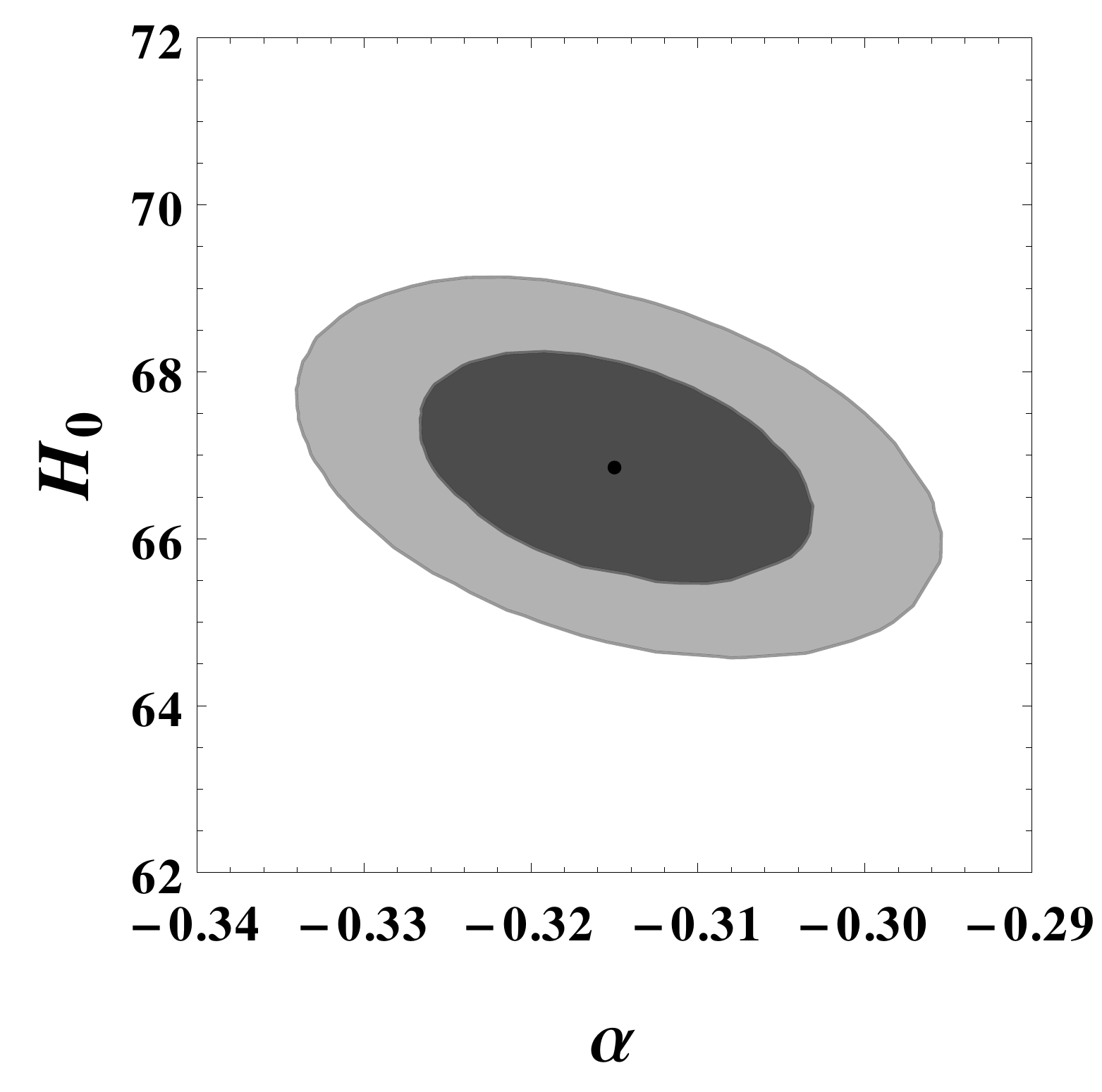}}
\end{center}
\caption{This figure shows the 1$\protect\sigma $ (dark shaded) and 2$\protect%
\sigma $ (light shaded) likelihood contours in $\protect\alpha -H_{0}$
plane. The figure corresponds to joint datasets of $H(z)+SN+BAO$. A black dot
represents the best-fit values of the model parameters which are found to be $
\protect\alpha =-0.3147$ and $H_{0}=66.84~Km/S/Mpc$.}
\label{fig:cont}
\end{figure}

\begin{figure}[tbph]
\begin{center}
\begin{tabular}{cc}
{\includegraphics[width=2.3in,height=2.3in,angle=0]{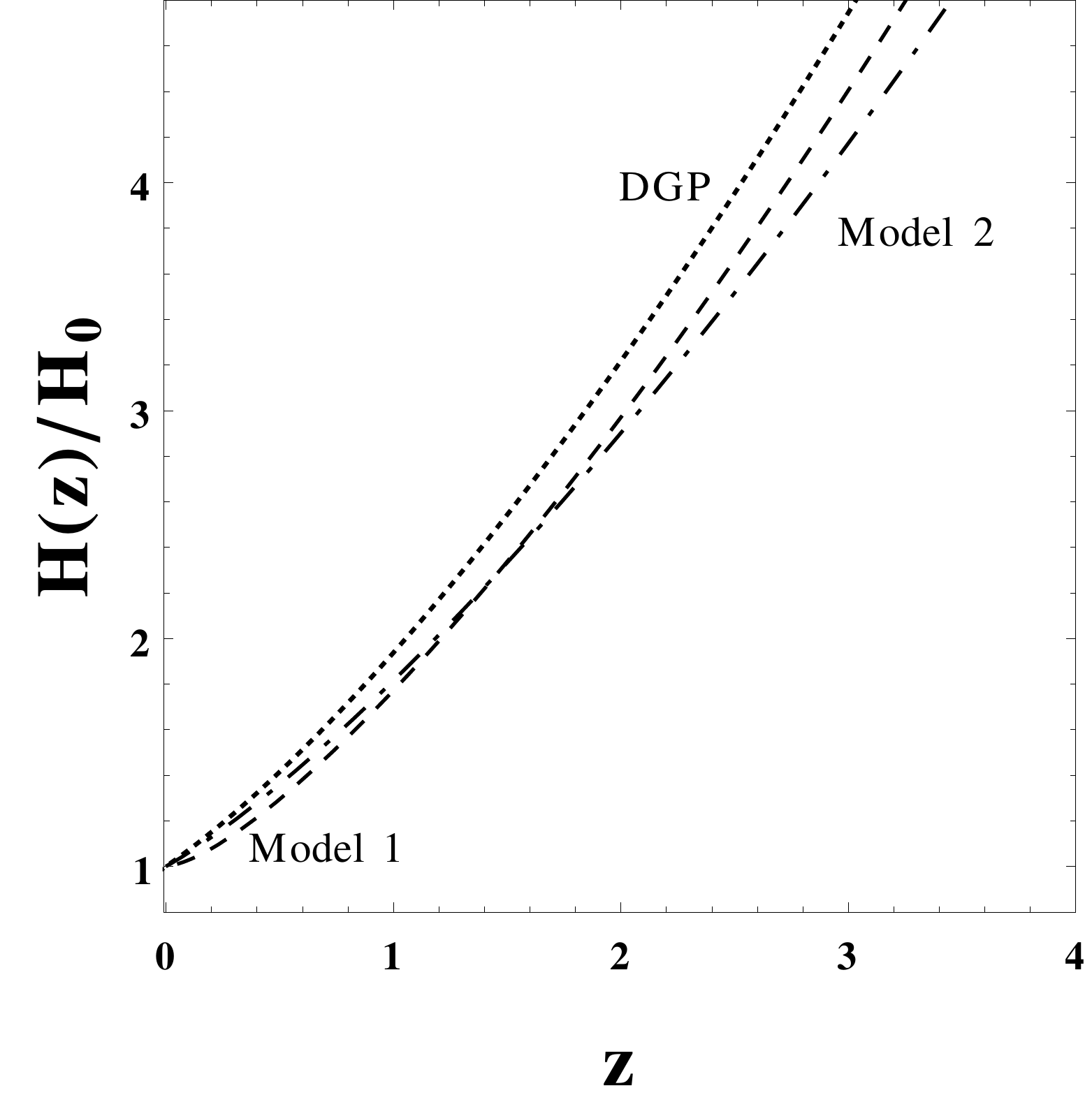}} & {%
\includegraphics[width=2.3in,height=2.3in,angle=0]{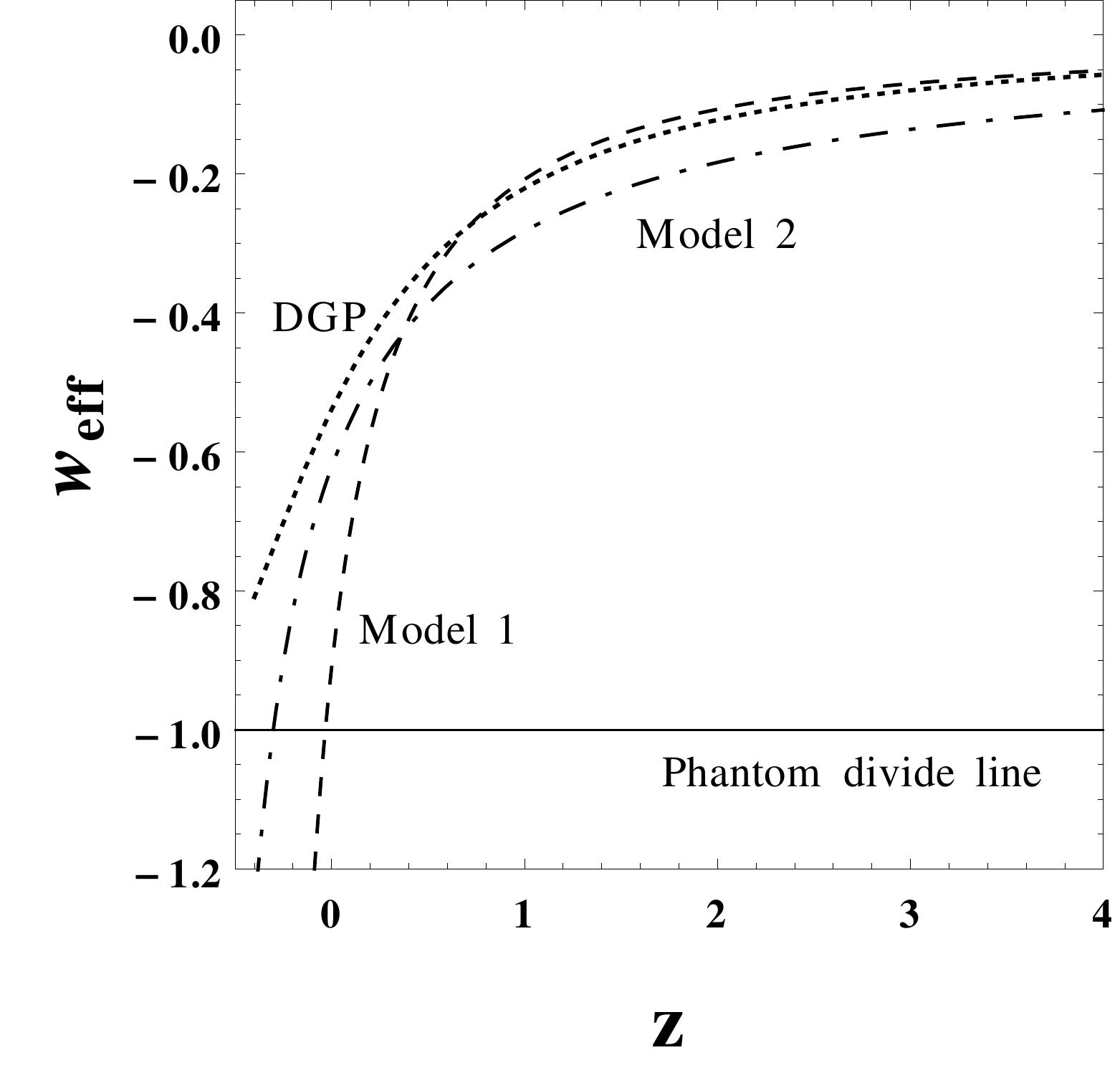}}%
\end{tabular}%
\end{center}
\caption{The figure shows the evolution of normalized Hubble parameter ($H(z)/H_0$)
and effective EOS ($w_{eff}$) versus redshift ($z$). The
dotted, dashed and dot-dashed lines correspond to DGP, models 1 and 2,
respectively. We use best-fit values of the model parameters. The horizontal line is the phantom divide line.}
\label{fig:hw}
\end{figure}

\begin{figure}[tbph]
\begin{center}
\begin{tabular}{cc}
{\includegraphics[width=2.3in,height=2.3in,angle=0]{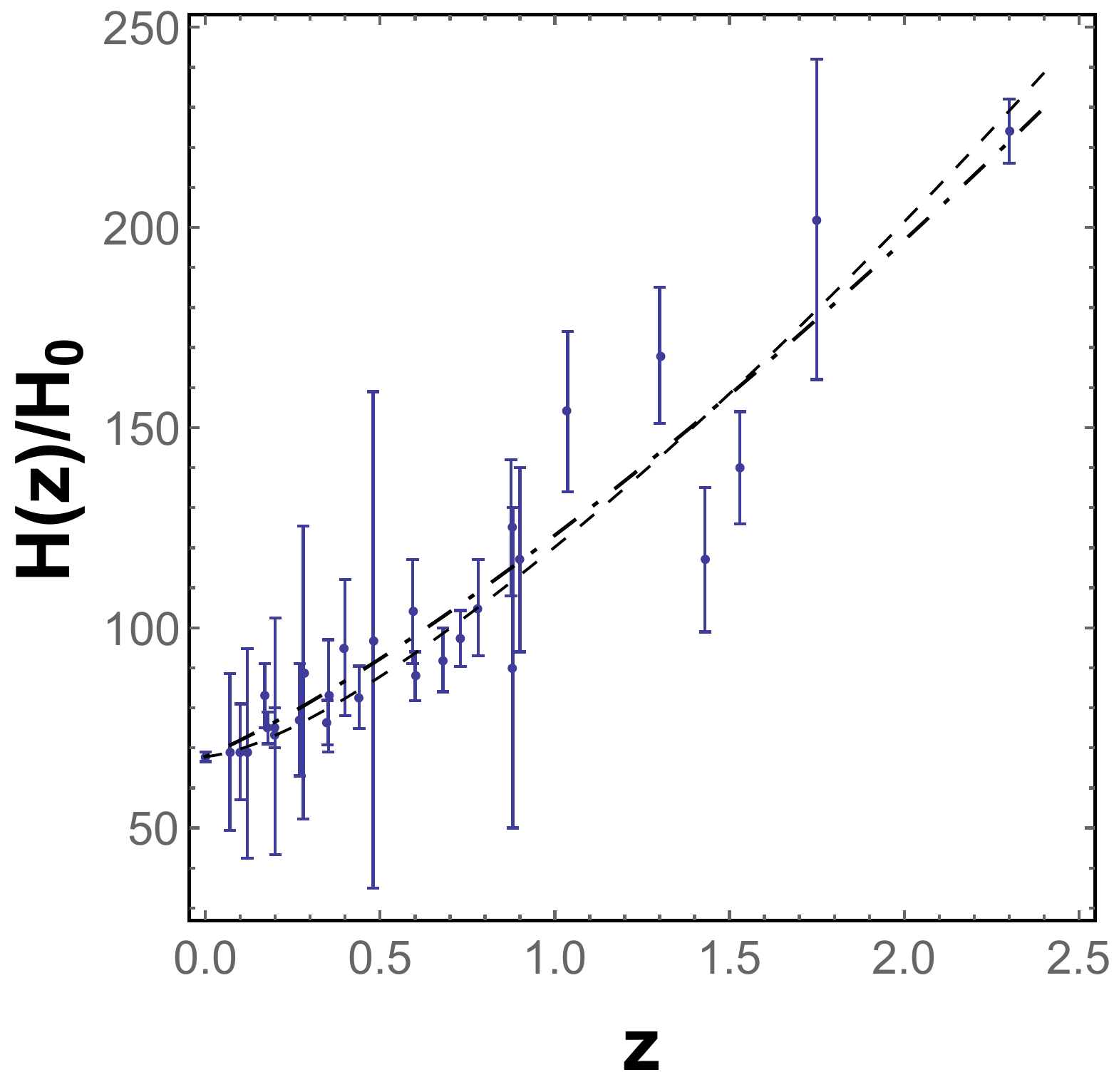}} & {%
\includegraphics[width=2.3in,height=2.3in,angle=0]{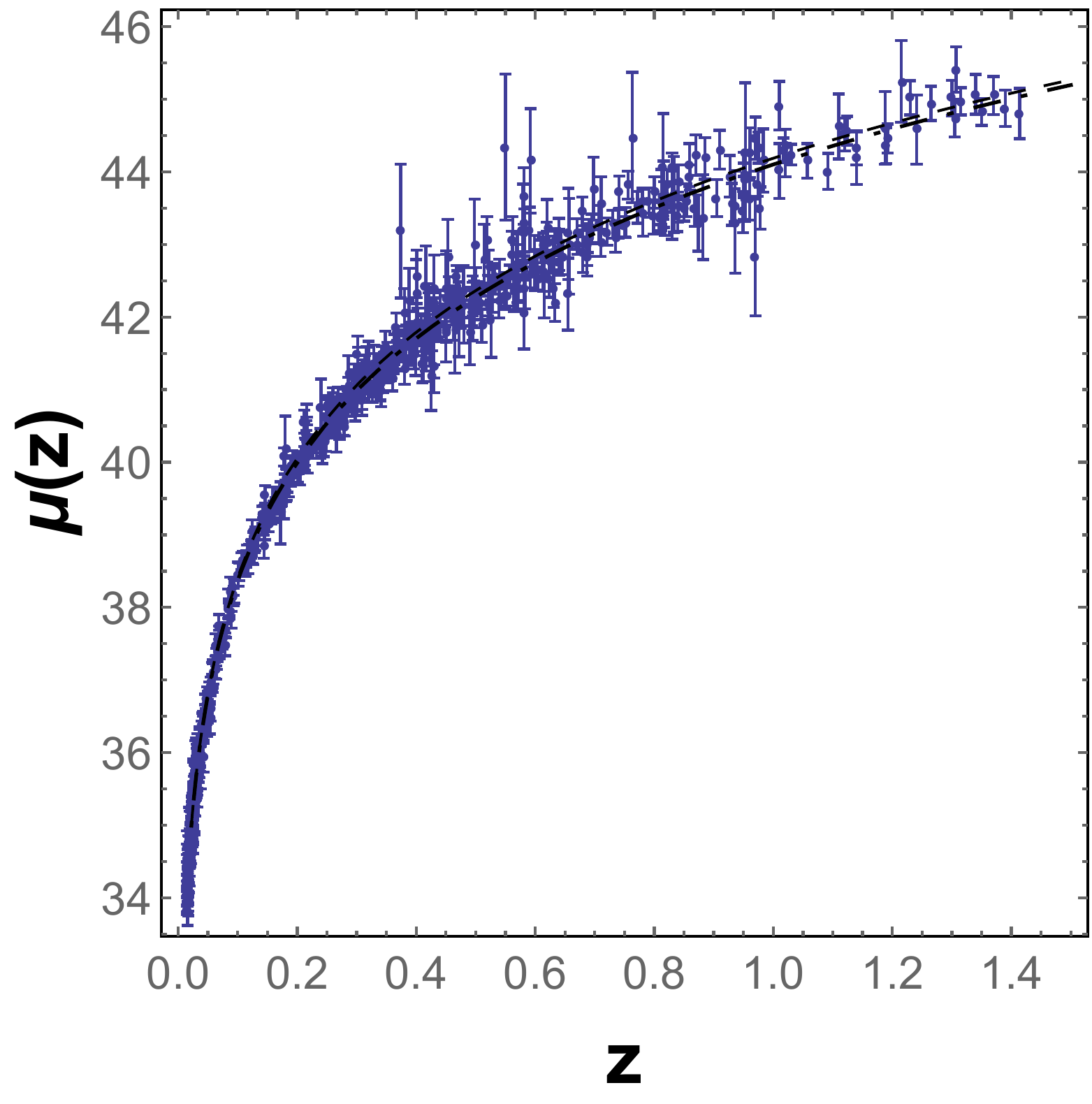}}%
\end{tabular}%
\end{center}
\caption{This figure exhibits the error bars of $H(z)$ (left) and $SN$ (right) datasets. In both the panels, the dashed and dot-dashed lines show the best-fitted behavior for models 1 and 2, respectively.}
\label{fig:mu}
\end{figure}

For model 2, we use an integrated datasets of $H(z)+SN+BAO$, and corresponding likelihood contour at 1$\sigma $ and 2$\sigma $ confidence levels are shown in Fig. \ref{fig:cont}. The best-fit values of the model parameters are obtained as $
\protect\alpha =-0.3147$ and $H_{0}=66.84~Km/S/Mpc$.
The normalized Hubble parameter and the effective EOS ($w_{eff}$) are plotted for both the models with their respective best-fit values of the parameters, and are shown in Fig. \ref{fig:hw}. Both the models represent phantom behavior in future. Model 1 exhibits the sudden singularity in near future while this kind of singularity has been delayed and pushed into far future in case of model 2 that is displayed in the right panel of Fig. \ref{fig:hw}. Fig. \ref{fig:mu} exhibits the error bar plots for models 1 and 2 with $H(z)$ and $SN$ datasets which shows that both the models are consistent with the observations.

The above discussions show the validation of our models corresponding to the
observations. In the following sections we shall employ different diagnostics for underlying models.
\begin{figure}[tbph]
\begin{center}
\begin{tabular}{cc}
{\includegraphics[width=2.3in,height=2.3in,angle=0]{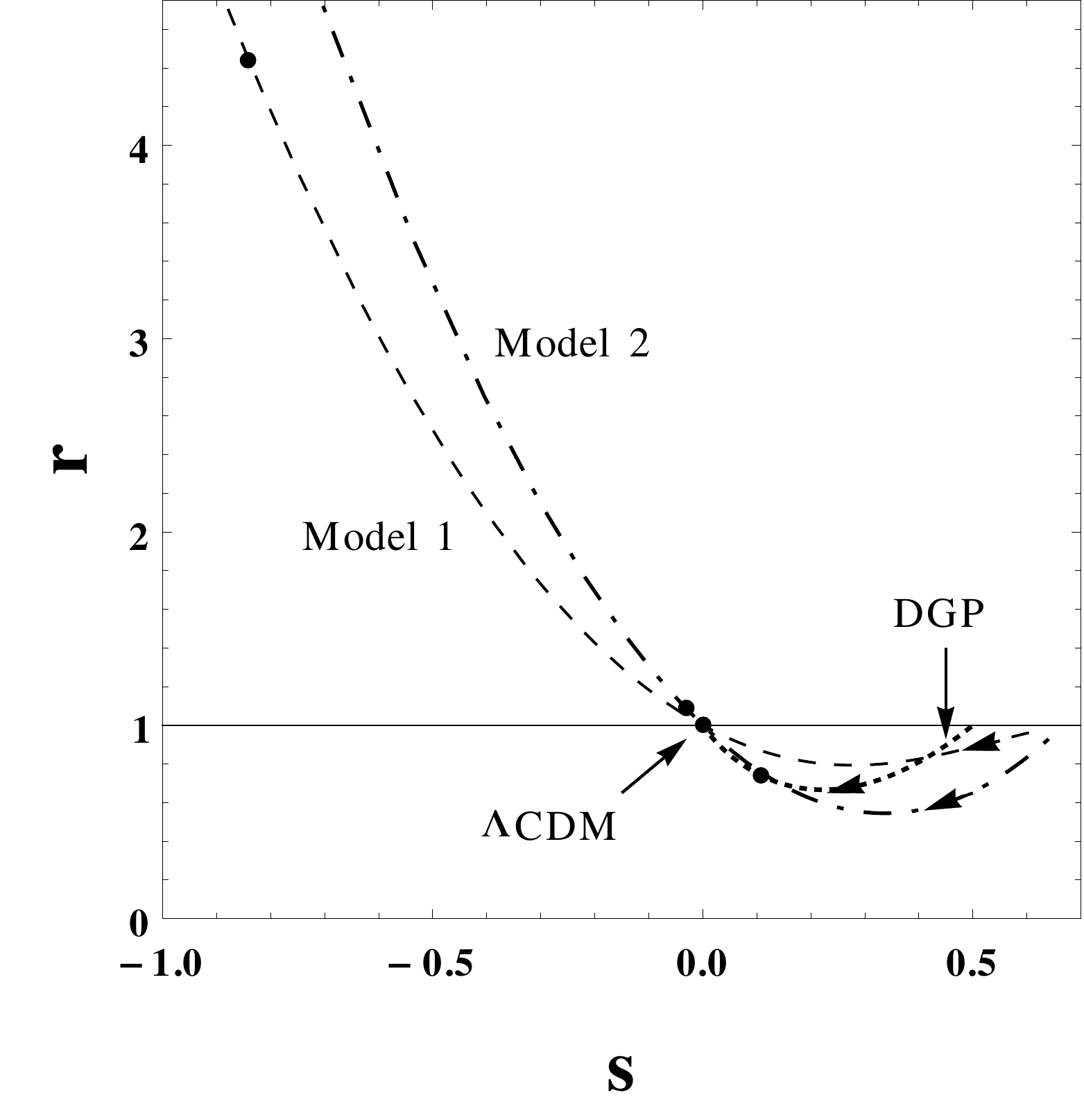}} & {%
\includegraphics[width=2.3in,height=2.3in,angle=0]{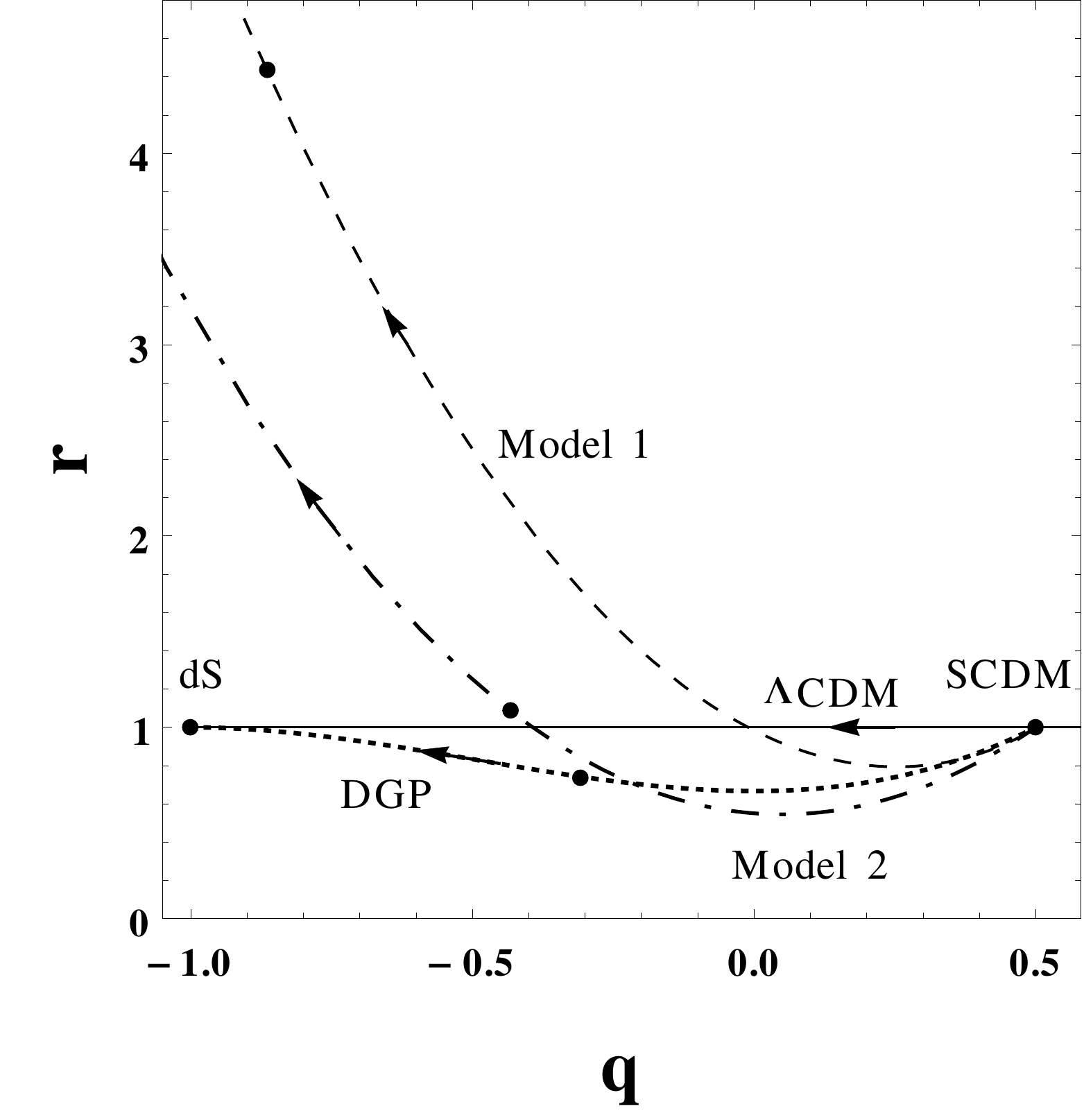}}%
\end{tabular}%
\end{center}
\caption{The figure shows the time evolution of the statefinder pairs $\{r,s\} $ (left) and $\{r,q\}$ (right) for models 1 (dashed), 2 (dot-dashed) and DGP (dotted). In the left panel, the fixed point ($r=1$, $s=0 $) corresponds to $\Lambda$CDM. All the models passes through $\Lambda$CDM. In the right panel, all models diverge from the same point ($r=1$, $q=0.5$) which corresponds to SCDM. The DGP converges to the
point ($r=1,q=-1$) that represents the de-Sitter expansion (dS) whereas models 1 and 2 do not converge to dS due to phantom nature. The dark dots on the curves denote current values $\{r_{0},s_{0}\}$ (left) and $\{r_{0},q_{0}\}$ (right). In all models,
we have taken best-fit values of the model parameters.}
\label{fig:rs}
\end{figure}
\section{Statefinder diagnostic}
\label{sec:state}
The past two decades produced a plethora of theoretical cosmological
models of dark energy with improved quality of observational data. So, there
should be some analysis which can differentiate these models and predict
the deviations from $\Lambda$CDM. Sahni et al. \cite{Sahni1} have pointed out this idea widely known as statefinder diagnostic. The statefinder pairs $\{r,s\}$ and $\{r,q\}$ are the geometrical quantities that are constructed from any space-time metric directly and can successfully differentiate various competing models of dark energy by using the higher order derivatives of scale factor. In the literature, the $\{r,s\}$ and $\{r,q\}$ pairs are defined as \cite{Sahni1}.
\begin{equation}
q=-\frac{\ddot{a}}{aH^{2}}\text{, \ \ } r=\frac{\dddot{a}}{aH^{3}}\text{, \ \ }s=\frac{r-1}{3(q-\frac{1}{2})}\text{.}
\label{d1}
\end{equation}%
The statefinder diagnostic is an useful tool in modern day cosmology and
being used to serve the purpose of distinguishing different dark energy
models \cite{alam1,alam2,alam3}. In this process, different
trajectories in $r-s$ and $r-q$ planes are plotted for various dark energy models
and study their behaviors. In a spatially flat FLRW background, the statefinder pair $\{r,s\}=\{1,0\}$ and $\{1,1\}$ for $\Lambda$CDM and standard cold dark matter (SCDM). In the $r-s$ and $r-q$ planes, the departure of any dark energy model from these fixed points are analyzed. The pairs $\{r,s\}$ and $\{r,q\}$ for models 1 and 2 are calculated as
\begin{equation}
r=\frac{%
\begin{array}{c}
11+101\alpha \tilde{a}+(358\alpha ^{2}+158\beta )\tilde{a}^{2}+(488\alpha
^{3}+1237\alpha \beta )\tilde{a}^{3} \\ 
+(2440\alpha ^{2}\beta +224\alpha ^{4}+1156\beta ^{2})\tilde{a}%
^{4}+(1424\alpha ^{3}\beta +4087\alpha \beta ^{2})\tilde{a}^{5} \\ 
+(3346\alpha ^{2}\beta ^{2}+2178\beta ^{3})\tilde{a}^{6}+3375\alpha \beta
^{3}\tilde{a}^{7}+1233\beta ^{4}\tilde{a}^{8}%
\end{array}%
}{2(1+\alpha \tilde{a}+\beta \tilde{a}^{2})^{2}(1+2\alpha \tilde{a}+3\beta 
\tilde{a}^{2})^{2}}  \label{d3}
\end{equation}%
\begin{equation}
s=\frac{%
\begin{array}{c}
9+89\alpha \tilde{a}+(332\alpha ^{2}+142\beta )\tilde{a}^{2}+(464\alpha
^{3}+1169\alpha \beta )\tilde{a}^{3} \\ 
+(216\alpha ^{4}+2348\alpha ^{2}\beta +1112\beta ^{2})\tilde{a}%
^{4}+(1384\alpha ^{3}\beta +3971\alpha \beta ^{2})\tilde{a}^{5} \\ 
+(3272\alpha ^{2}\beta ^{2}+2130\beta ^{3})\tilde{a}^{6}+3315\alpha \beta
^{3}\tilde{a}^{7}+1215\beta ^{4}\tilde{a}^{8}%
\end{array}%
}{%
\begin{array}{c}
15\alpha \tilde{a}+(63\alpha ^{2}+42\beta )\tilde{a}^{2}+(84\alpha
^{3}+255\alpha \beta )\tilde{a}^{3} \\ 
+(36\alpha ^{4}+438\alpha ^{2}\beta +222\beta ^{2})\tilde{a}^{4}+(228\alpha
^{3}\beta +693\alpha \beta ^{2})\tilde{a}^{5} \\ 
+(507\alpha ^{2}\beta ^{2}+342\beta ^{3})\tilde{a}^{6}+477\alpha \beta ^{3}%
\tilde{a}^{7}+162\beta ^{4}\tilde{a}^{8}%
\end{array}%
}  \label{d4}
\end{equation}
and
\begin{equation}
r=\frac{3+(2+18\alpha )\tilde{a}+(4\alpha +41\alpha ^{2})\tilde{a}%
^{2}+(2\alpha ^{2}+33\alpha ^{3})\tilde{a}^{3}+9\alpha ^{4}\tilde{a}^{4}}{2%
\tilde{a}+4\alpha \tilde{a}^{2}+2\alpha ^{2}\tilde{a}^{3}}  \label{d5}
\end{equation}%
\begin{equation}
s=\frac{-3-18\alpha \tilde{a}-41\alpha ^{2}\tilde{a}^{2}-33\alpha ^{3}\tilde{%
a}^{3}-9\alpha ^{4}\tilde{a}^{4}}{-9+(9-30\alpha )\tilde{a}+(18\alpha
-30\alpha ^{2})\tilde{a}^{2}+(9\alpha ^{2}-9\alpha ^{3})\tilde{a}^{3}}
\label{d6}
\end{equation}
The deceleration parameter $q$ is given by Eqs. (\ref{eq:q1}) and (\ref{eq:q2}), respectively. We plot the $r-s$ and $r-q$ diagrams for our models and compare these with the $\Lambda$CDM.

Fig. \ref{fig:rs} shows the time evolution of the statefinder pairs for different DE models. The left panel exhibits the evolution of $\{r,s\}$ while the right one for $\{r,q\}$. In both panels, the models 1 (dashed) and 2 (dot-dashed) are compared with DGP (dotted) and $\Lambda$CDM. In left panel, the fixed point ($r=1$, $s=0$)
corresponds to $\Lambda$CDM, and all models passes through this fixed point. Moreover, one can see that the trajectory of DGP terminates there while the corresponding
trajectories of models 1 and 2 evolve further showing that the phantom behavior in future. In right panel, the fixed point ($r=1$, $q=0.5$) represents SCDM. All the underlying models evolve from this point. The DGP converges to the second point ($r=1,q=-1$) that corresponds to de-Sitter expansion (dS) whereas models 1 and 2 do not
converge to dS due to their phantom behavior. The dark dots
on the curves show present values $\{r_{0},s_{0}\}$ (left) and $%
\{r_{0},q_{0}\}$ (right) for the models under consideration. We chose the best-fit values of the model parameters for these plots.
\section{$Om$ diagnostic}
\label{sec:om}
We shall now diagnose our models with $Om$ analysis which
is also a geometrical diagnostic that explicitly depends on redshift and the Hubble
parameter, and is defined as \cite{Sahni2,Z}:
\begin{equation}
Om\left( z\right) =\frac{\left( \frac{H(z)}{H_{0}}\right)^{2}-1}{\left( 1+z\right)
^{3}-1}\text{} 
\label{eq:om}
\end{equation}
\begin{figure}[tbph]
\begin{center}
{\includegraphics[width=2.5in,height=2.5in,angle=0]{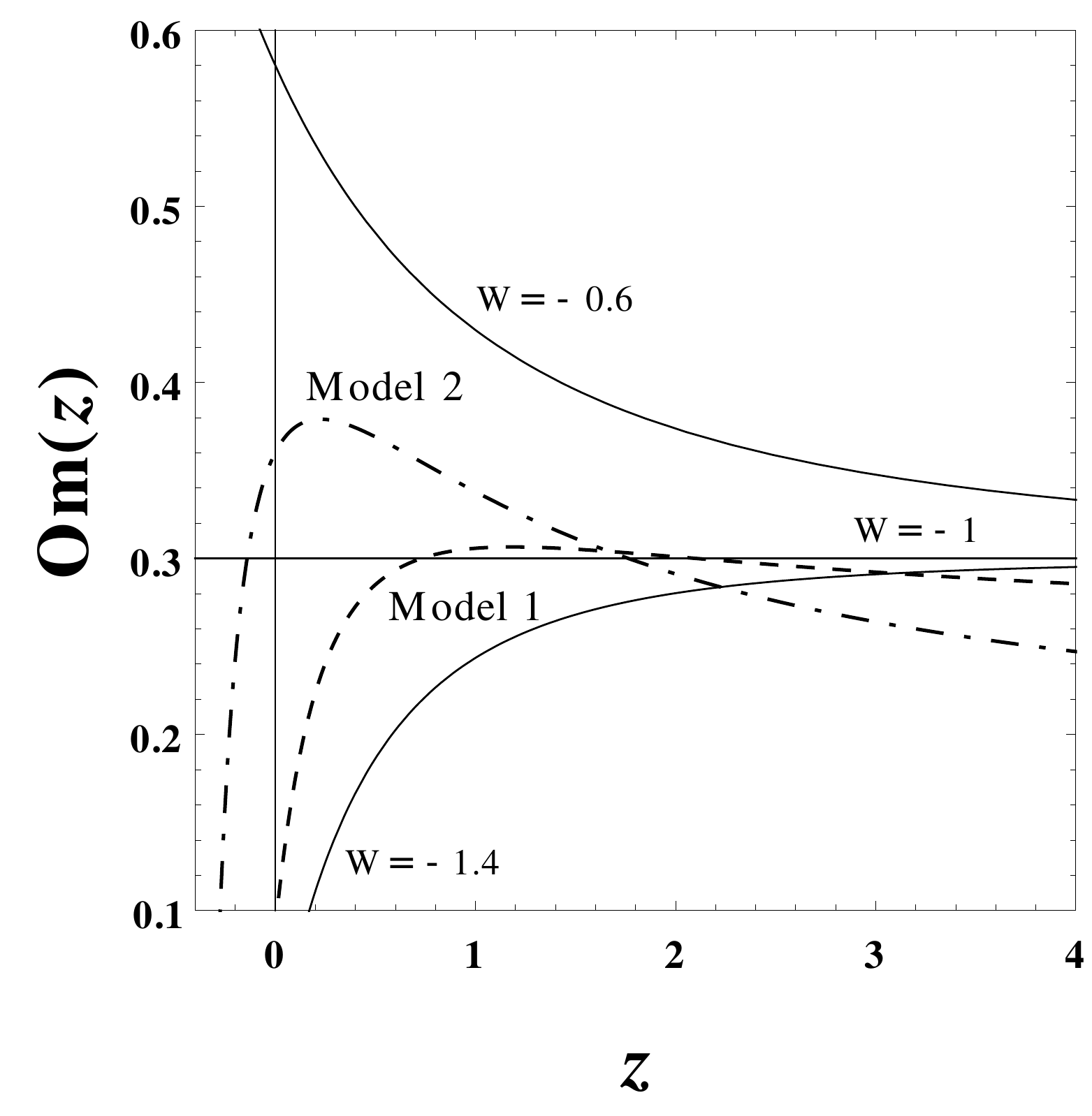}}
\end{center}
\caption{This figure shows the evolution of $Om(z)$ versus redshift $z$
for different DE models such as $w=-1$ ($\Lambda$CDM), $-1.4$ (phantom), $-0.6$ (quintessence) and models 1 (dashed) \& 2 (dot-dashed). The horizontal line represents $\Lambda$CDM, and has zero curvature. The DE models with $w>-1$
(quintessence) have negative curvature whereas models with $w<-1$
(phantom) have positive curvature. Models 1 and 2 show the positive curvatures though they have $w_{eff}>-1$ at the present epoch and in future $w_{eff}<-1$
(phantom phase). The vertical solid line represents the present era. The
best-fit values are chosen to plot this figure.}
\label{fig:om}
\end{figure}
The $Om$ diagnostic also differentiates various DE models from $\Lambda$CDM \cite{Amnras}. This is a simpler diagnostic when applied to observations as it depends only on the first derivative of scale factor. The Hubble parameter for a constant EOS is defined as
\begin{equation}
H^2(z) = H_0^2 \left[ \Omega_{0m}(1+z)^3 + (1-\Omega_{0m})(1+z)^{3(1+w)}\right] ,
\label{eq:Hconst}
\end{equation}
The expression for $Om(z)$ corresponding to Eq. (\ref{eq:Hconst}) is written as
\begin{equation}
Om(z) = \Omega_{0m} + (1-\Omega_{0m})\frac{(1+z)^{3(1+w)}-
1}{(1+z)^3-1}
\label{eq:omconst}
\end{equation} 
From Eq. (\ref{eq:omconst}), one can notice that, $Om(z) = \Omega_{0m}$ for $w=-1$ ($\Lambda$CDM) whereas $Om(z) > \Omega_{0m}$ for $w> -1$ (quintessence) and $Om(z) < \Omega_{0m}$ for $w < -1$ (phantom). The corresponding evolutions of $Om(z)$ for $w=-0.6$ (quintessence), $-1$ ($\Lambda$CDM) and $-1.4$ (phantom) are shown in Fig. \ref{fig:om}. From Fig. \ref{fig:om}, one can clearly see that the $Om(z)$ has negative, zero and positive curvatures for quintessence, $\Lambda$CDM and phantom, respectively. In contrast, we show the evolution of $Om(z)$ for models 1 and 2 in Fig. \ref{fig:om}. Both models exhibit positive curvatures though thay have $w_{eff}>-1$ (quintessence) at present epoch and in future $w_{eff}<-1$ (phantom phase), see Figs. \ref{fig:hw} and \ref{fig:om}. This is a vital result as in the literature, quintessence does not have positive curvature which is a generic feature \cite{Amnras}. 
\section{Conclusion}
\label{sec:conc}
In this work, we have considered the scenario in which cosmic
acceleration might arise due to coupling between known matter components
present in the Universe \cite{KHOURY2016}. To this effect, we further extend
the work of Ref. \cite{ARSSA}. The two models obtained by parameterizing the
coupling function (or correspondingly the Einstein frame scale factor in
terms of physical scale factor) are analyzed by using an integrated observational data.
We used joint data of $H(z)$, Type Ia supernova and BAO, and constrained the model parameters of model 2 (see Fig. \ref{fig:cont}). In this case, the best-fit values of 
$\alpha$ and $H_0$ are found to be $\alpha =-0.3147$ and $H_{0}=66.84~Km/S/Mpc$ (model 1 consists of two parameters that were constrained in Ref. \cite{ARSSA} and the best-fit values were obtained as $\alpha =-0.102681$ \& $\beta =-0.078347$). We used best-fit values of the model parameters to carried out the analysis. The time evolutions of Hubble parameter and effective EOS  are shown in Fig. \ref{fig:hw}. From this figure we conclude that the model 1 shows phantom behavior with a pressure singularity in the near future while the model 2 is a generalized case of model 1 that pushes the future sudden singularity to the infinite future. In Fig. \ref{fig:mu}, we have shown error bars of observational data with the models under consideration. One can clearly observe that both the models are compatible with the observations.

In addition, the statefinder diagnostic has been performed for the underlying models. We obtained the expressions for the statefinder pairs and their behaviors have been displayed in $r-s$ and $r-q$ planes as shown in Fig. \ref{fig:rs}. For comparison, we have also shown DGP model in same figure. In the $r-s$ plane, both the models pass
through the fixed point ($r=1, s=0$) and move away from the $\Lambda$CDM while the
DGP terminates at the fixed point. This is due the phantom behavior which is a distinguished characteristic of the underlying scenario. In the $q-r$ plane, it is clearly seen that all the models originated from a
fixed point ($r=1,q=0.5$) that corresponds to SCDM.
The DGP model converges to the second point ($r=1,q=-1$) that represents 
the dS while the models under consideration do not
converge to the dS fixed point due to their phantom nature (see right
panel of Fig. \ref{fig:rs}). 

The evolution of $Om(z)$ versus redshift $z$ for different DE models are shown in Fig. \ref{fig:om}. We observed that the $\Lambda$CDM, quintessence and phantom have zero, negative and positive curvatures. Models 1 and 2 lie in quintessence regime in the past and remain so till the present epoch, and evolve to phantom in future. Both the models have positive curvature however they lie in the quintessence regime. This is an important result as in the literature, its not possible to have positive curvature for quintessence \cite{Amnras}.

In our opinion, the scenario proposed in \cite{KHOURY2016} and investigated
here is of great interest in view of GW170817. The modification caused by
the interaction between known components of matter does not involve any
extra degree of freedom and falls into the safe category in the light of
recent observations on gravitational waves.

\section*{Acknowledgment}
We are highly indebted to M. Sami for suggesting this problem and constant supervision as well as for providing all necessities to complete this work. SKJP wishes to
thank National Board of Higher Mathematics (NBHM), Department of Atomic
Energy (DAE), Govt. of India for financial support through the post-doctoral
research fellowship.


\begin{thebibliography}{99}
\bibitem{HZTEAM} A.~G.~Riess \textit{et al.} [Supernova Search Team], 
Astron.\ J.\ \textbf{116}, 1009 (1998) 
[astro-ph/9805201]. 

\bibitem{SCP} S.~Perlmutter \textit{et al.} [Supernova Cosmology Project
Collaboration], 
Astrophys.\ J.\ \textbf{517}, 565 (1999) 
[astro-ph/9812133]. 

\bibitem{cmb1} A.~H.~Jaffe \textit{et al.} [Boomerang Collaboration], 
Phys.\ Rev.\ Lett.\ \textbf{86}, 3475 (2001) 
[astro-ph/0007333].

\bibitem{cmb2} D.~N.~Spergel \textit{et al.} [WMAP Collaboration], 
Astrophys.\ J.\ Suppl.\ \textbf{170}, 377 (2007) 
[astro-ph/0603449].

\bibitem{bao1} J.~R.~Bond, G.~Efstathiou and M.~Tegmark, 
Mon.\ Not.\ Roy.\ Astron.\ Soc.\ \textbf{291}, L33 (1997) 
[astro-ph/9702100].

\bibitem{bao2} Y.~Wang and P.~Mukherjee, 
Astrophys.\ J.\ \textbf{650}, 1 (2006) 
[astro-ph/0604051].

\bibitem{sdss1} U.~Seljak \textit{et al.} [SDSS Collaboration], 
Phys.\ Rev.\ D \textbf{71}, 103515 (2005) 
[astro-ph/0407372]. 

\bibitem{sdss2} J.~K.~Adelman-McCarthy \textit{et al.} [SDSS Collaboration], 
Astrophys.\ J.\ Suppl.\ \textbf{162}, 38 (2006) 
[astro-ph/0507711].

\bibitem{DE1} E.~J.~Copeland, M.~Sami and S.~Tsujikawa, 
Int.\ J.\ Mod.\ Phys.\ D \textbf{15}, 1753 (2006) 
[hep-th/0603057]. 

\bibitem{DE2} M.~Sami, 
New Adv.\ Phys.\ \textbf{10}, 77 (2016) [arXiv:1401.7310 [physics.pop-ph]]. 

\bibitem{DE3} V.~Sahni and A.~A.~Starobinsky, 
Int.\ J.\ Mod.\ Phys.\ D \textbf{9}, 373 (2000) 
[astro-ph/9904398]. 

\bibitem{DE4} J.~Frieman, M.~Turner and D.~Huterer, 
Ann.\ Rev.\ Astron.\ Astrophys.\ \textbf{46}, 385 (2008) 
[arXiv:0803.0982 [astro-ph]]. 
%

\bibitem{DE5} R.~R.~Caldwell and M.~Kamionkowski, 
Ann.\ Rev.\ Nucl.\ Part.\ Sci.\ \textbf{59}, 397 (2009) 
[arXiv:0903.0866 [astro-ph.CO]]. 

\bibitem{DE6} A.~Silvestri and M.~Trodden, 
Rept.\ Prog.\ Phys.\ \textbf{72}, 096901 (2009) 
[arXiv:0904.0024 [astro-ph.CO]]. 

\bibitem{DE7} M.~Sami, 
Curr.\ Sci.\ \textbf{97}, 887 (2009) [arXiv:0904.3445 [hep-th]]. 

\bibitem{DE8} L.~Perivolaropoulos, 
AIP Conf.\ Proc.\ \textbf{848}, 698 (2006) 
[astro-ph/0601014]. 

\bibitem{DE9} J.~A.~Frieman, 
AIP Conf.\ Proc.\ \textbf{1057}, 87 (2008) 
[arXiv:0904.1832 [astro-ph.CO]]. 

\bibitem{DE10} M.~Sami, 
Lect.\ Notes Phys.\ \textbf{720}, 219 (2007). 

\bibitem{carollCC} S.~M.~Carroll, 
Living Rev.\ Rel.\ \textbf{4}, 1 (2001) 
[astro-ph/0004075]. 

\bibitem{padmaCC} T.~Padmanabhan, 
Phys.\ Rept.\ \textbf{380}, 235 (2003) 
[hep-th/0212290]. 

\bibitem{peeblesCC} P.~J.~E.~Peebles and B.~Ratra, 
Rev.\ Mod.\ Phys.\ \textbf{75}, 559 (2003) 
[astro-ph/0207347]. 

\bibitem{quint1} C.~Wetterich, 
Nucl.\ Phys.\ B \textbf{302}, 668 (1988). 

\bibitem{quint2} B.~Ratra and P.~J.~E.~Peebles, 
Phys.\ Rev.\ D \textbf{37}, 3406 (1988). 

\bibitem{quint3} R.~R.~Caldwell, R.~Dave and P.~J.~Steinhardt, 
Phys.\ Rev.\ Lett.\ \textbf{80}, 1582 (1998) 
[astro-ph/9708069]. 

\bibitem{quint4} V.~Sahni, M.~Sami and T.~Souradeep, 
Phys.\ Rev.\ D \textbf{65}, 023518 (2002) 
[gr-qc/0105121]. 

\bibitem{quint5} M.~Sami and T.~Padmanabhan, 
Phys.\ Rev.\ D \textbf{67}, 083509 (2003) 
[hep-th/0212317]. 
[hep-th]]. 

\bibitem{phant1} L.~Parker and A.~Raval, 
Phys.\ Rev.\ D \textbf{60}, 063512 (1999) 
[gr-qc/9905031].

\bibitem{phant2} V.~Sahni and A.~A.~Starobinsky, 
Int.\ J.\ Mod.\ Phys.\ D \textbf{9}, 373 (2000) 
[astro-ph/9904398].

\bibitem{phant3} S.~Nojiri and S.~D.~Odintsov, 
Phys.\ Lett.\ B \textbf{562}, 147 (2003) 
[hep-th/0303117].

\bibitem{phant4} P.~Singh, M.~Sami and N.~Dadhich, 
Phys.\ Rev.\ D \textbf{68}, 023522 (2003) 
[hep-th/0305110].

\bibitem{phant5} M.~Sami and A.~Toporensky, 
Mod.\ Phys.\ Lett.\ A \textbf{19}, 1509 (2004) 
[gr-qc/0312009].

\bibitem{phant6} M.~Sami, A.~Toporensky, P.~V.~Tretjakov and S.~Tsujikawa, 
Phys.\ Lett.\ B \textbf{619}, 193 (2005) 
[hep-th/0504154].

\bibitem{tachy1} A.~Sen, 
JHEP \textbf{0207}, 065 (2002) 
[hep-th/0203265].

\bibitem{tachy2} T.~Padmanabhan, 
Phys.\ Rev.\ D \textbf{66}, 021301 (2002) 
[hep-th/0204150].

\bibitem{alam2017} M. Shahalam, S.D. Pathak, Shiyuan Li, R. Myrzakulov, Anzhong Wang, Eur. Phys. J. C 77 (2017) 686. 

\bibitem{CG1} A.~Y.~Kamenshchik, U.~Moschella and V.~Pasquier, 
Phys.\ Lett.\ B \textbf{511}, 265 (2001) 
[gr-qc/0103004].

\bibitem{CG2} V.~Gorini, U.~Moschella, A.~Kamenshchik and V.~Pasquier, 
AIP Conf.\ Proc.\ \textbf{751}, 108 (2005). 

\bibitem{DEREV1} K.~Bamba, S.~Capozziello, S.~Nojiri and S.~D.~Odintsov, 
Astrophys.\ Space Sci.\ \textbf{342}, 155 (2012) 
[arXiv:1205.3421 [gr-qc]].

\bibitem{DEREV3} A.~Ali, R.~Gannouji and M.~Sami, 
Phys.\ Rev.\ D \textbf{82}, 103015 (2010) 
[arXiv:1008.1588 [astro-ph.CO]].

\bibitem{DEREV2} J.~Yoo and Y.~Watanabe, 
Int.\ J.\ Mod.\ Phys.\ D \textbf{21}, 1230002 (2012) 
[arXiv:1212.4726 [astro-ph.CO]].

\bibitem{KHOURY2016} L.~Berezhiani, J.~Khoury and J.~Wang, 
Phys.\ Rev.\ D \textbf{95}, no. 12, 123530 (2017) 
[arXiv:1612.00453 [hep-th]]. 

\bibitem{ARSSA} A.~Agarwal, R.~Myrzakulov, S.~K.~J.~Pacif, M.~Sami and
A.~Wang, 
arXiv:1709.02133 [gr-qc].

\bibitem{dgp}
G. Dvali, G. Gabadadze and M. Porrati, 4D Gravity on a Brane in 5D Minkowski Space, Phys. Lett. B {\bf 485}, 208 (2000).

\bibitem{Farooq:2013hq} O.~Farooq and B.~Ratra, 
Astrophys.\ J.\ \textbf{766}, L7 (2013) [arXiv:1301.5243 [astro-ph.CO]]. And
the references their in.

\bibitem{planck} P.~A.~R.~Ade \textit{et al.} [Planck Collaboration], A \&
A, 571, A16 (2014).

\bibitem{planck2015} P.~A.~R.~Ade \textit{et al.} [Planck Collaboration], A \&
A, 594, A13 (2016).

\bibitem{Suzuki:2011hu} N.~Suzuki, D.~Rubin, C.~Lidman, G.~Aldering,
R.~Amanullah, K.~Barbary, L.~F.~Barrientos and J.~Botyanszki \textit{et al.}%
, 
Astrophys.\ J.\ \textbf{746}, 85 (2012) [arXiv:1105.3470 [astro-ph.CO]]. 

\bibitem{Giostri:2012ek} R.~Giostri, M.~V.~d.~Santos, I.~Waga,
R.~R.~R.~Reis, M.~O.~Calvao and B.~L.~Lago, 
JCAP \textbf{1203}, 027 (2012) [arXiv:1203.3213 [astro-ph.CO]]. 
%
%
%
%
%

\bibitem{Blake:2011en} C.~Blake, E.~Kazin, F.~Beutler, T.~Davis,
D.~Parkinson, S.~Brough, M.~Colless and C.~Contreras \textit{et al.}, 
Mon.\ Not.\ Roy.\ Astron.\ Soc.\ \textbf{418}, 1707 (2011) [arXiv:1108.2635
[astro-ph.CO]]. 


\bibitem{Percival:2009xn} W.~J.~Percival \textit{et al.} [SDSS
Collaboration], 
Mon.\ Not.\ Roy.\ Astron.\ Soc.\ \textbf{401}, 2148 (2010) [arXiv:0907.1660
[astro-ph.CO]]. 


\bibitem{Beutler:2011hx} F.~Beutler, C.~Blake, M.~Colless, D.~H.~Jones,
L.~Staveley-Smith, L.~Campbell, Q.~Parker and W.~Saunders \textit{et al.}, 
Mon.\ Not.\ Roy.\ Astron.\ Soc.\ \textbf{416}, 3017 (2011) [arXiv:1106.3366
[astro-ph.CO]]. 


\bibitem{Jarosik:2010iu} N.~Jarosik, C.~L.~Bennett, J.~Dunkley, B.~Gold,
M.~R.~Greason, M.~Halpern, R.~S.~Hill and G.~Hinshaw \textit{et al.}, 
Astrophys.\ J.\ Suppl.\ \textbf{192}, 14 (2011) [arXiv:1001.4744
[astro-ph.CO]]. 


\bibitem{Eisenstein:2005su} D.~J.~Eisenstein \textit{et al.} [SDSS
Collaboration], 
Astrophys.\ J.\ \textbf{633}, 560 (2005) [astro-ph/0501171]. 


\bibitem{Sahni1} V.~Sahni, T.~D.~Saini, A.~A.~Starobinsky and U.~Alam, 
JETP Lett.\ \textbf{77}, 201 (2003); U. Alam, V. Sahni, T. D. Saini,
and A. A. Starobinsky, Mon. Not. R. Astron. Soc. {\bf344}, 1057 (2003).

\bibitem{alam1} M. Sami {\it et al}., Cosmological dynamics of non-minimally coupled scalar field system and its late time cosmic relevance, Phys. Rev. D {\bf86} (2012) 103532 [arXiv:1207.6691] [ INSPIRE ].

\bibitem{alam2} R. Myrzakulov and M. Shahalam, Statefinder hierarchy of bimetric and galileon models for concordance cosmology, JCAP {\bf10} (2013) 047 [arXiv:1303.0194] [ INSPIRE ].

\bibitem{alam3}Sarita Rani {\it et al}., Constraints on cosmological parameters in power-law cosmology, JCAP {\bf 03} (2015) 031.

\bibitem{Sahni2} V.~Sahni, A.~Shafieloo and A.~A.~Starobinsky, 
Phys.\ Rev.\ D \textbf{78}, 103502 (2008) 
[arXiv:0807.3548 [astro-ph]].

\bibitem{Z} C.~Zunckel and C.~Clarkson, 
Phys.\ Rev.\ Lett.\ \textbf{101}, 181301 (2008) 
[arXiv:0807.4304 [astro-ph]].

\bibitem{Amnras} M. Shahalam, Sasha Sami, Abhineet Agarwal, $Om$ diagnostic applied to scalar field models and slowing down of cosmic acceleration, Mon. Not. Roy. Astron. Soc. {\bf448} (2015) 2948-2959 [arXiv:1501.04047] [astro-ph.CO]
\end{thebibliography}
\end{document}